# Enhancement and Inhibition of Spontaneous Photon Emission by Resonant Silicon Nanoantennas

Dorian Bouchet [1], Mathieu Mivelle [1,2], Julien Proust[3], Bruno Gallas[4], Igor Ozerov[5], Maria F. Garcia-Parajo [2,6], Angelo Gulinatti[7], Ivan Rech[7], Yannick De Wilde[1], Nicolas Bonod[3], Valentina Krachmalnicoff[1,†] and Sébastien Bidault[1,‡]

[1]*ESPCI Paris, PSL Research University, CNRS, Institut Langevin, 1 rue Jussieu, F-75005, Paris, France*
[2]*ICFO-Institut de Ciencies Fotoniques, The Barcelona Institute of Science and Technology, 08860 Castelldefels (Barcelona), Spain*
[3]*Aix Marseille Univ, CNRS, Centrale Marseille, Institut Fresnel, Marseille, France*
[4]*Sorbonne Universités, UPMC Univ Paris 06, CNRS UMR 7588, Institut des Nanosciences de Paris, 75005 Paris, France*
[5]*Aix Marseille Univ, CNRS, CINAM, Marseille, France*
[6]*ICREA-Institució Catalana de Recerca i Estudis Avançats, 08010 Barcelona, Spain*
[7]*Politecnico di Milano, Dipartimento di Elettronica, Informazione e Bioingegneria, Piazza da Vinci 32, 20133 Milano, Italy*
(Dated: September 14, 2016)

Abstract

Substituting noble metals for high-index dielectrics has recently been proposed as an alternative strategy in nanophotonics to design broadband optical resonators and circumvent the ohmic losses of plasmonic materials. In this report, we demonstrate that subwavelength silicon nanoantennas can manipulate the photon emission dynamics of fluorescent molecules. In practice, it is showed that dielectric nanoantennas can both increase and decrease the local density of optical states (LDOS) at room temperature, a process that is inaccessible with noble metals at the nanoscale. Using scanning probe microscopy, we analyze quantitatively, in three dimensions, the near-field interaction between a 100 nm fluorescent nanosphere and silicon nanoantennas with diameters ranging between 170 nm and 250 nm. Associated to numerical simulations, these measurements indicate increased or decreased total spontaneous decay rates by up to 15 % and a gain in the collection efficiency of emitted photons by up to 85 %. Our study demonstrates the potential of silicon-based nanoantennas for the low-loss manipulation of solid-state emitters at the nanoscale and at room temperature.

## I. INTRODUCTION

Broadband sub-wavelength optical resonators, or nanoantennas, have the ability to enhance the luminescence decay rate and brightness of solid-state emitters at room-temperature by increasing the local density of optical states (LDOS) [1-8]. Generally based on noble metal nanoparticles exhibiting surface plasmon polaritons, these resonators feature ohmic losses that locally quench spontaneous photon emission and reduce fluorescence lifetimes [2, 3, 8]. Importantly, these non-radiative processes prevent plasmonic antennas from inhibiting spontaneous emission in their vicinity as the LDOS can never be lower than the one found in vacuum. Even when plasmonic nanoantennas feature a dark mode weakly coupled to the far-field, non-radiative decay channels dominate and increase the decay rate of a coupled emitter [9, 10]. In contrast, high-quality factor dielectric cavities can exhibit a reduced LDOS to lengthen the excited state lifetime of narrowband emitters such as isolated atoms [11] or solid-state light sources at cryogenic temperatures [12-15].

High-index dielectric nanoparticles, featuring broadband low-order Mie resonances, have been recently proposed as alternatives for the design of optical nanoantennas with weak ohmic losses and strong magnetic modes [16-24], as well as enhanced local fields [25-29]. In particular, several theoretical studies have proposed to use subwavelength-sized silicon resonators to accelerate spontaneous decay processes from isolated emitters [30-33]. In this work, we experimentally demonstrate that silicon nanoantennas can be used to resonantly enhance but also to inhibit spontaneous photon emission in fluorescent molecules at room temperature thanks to their low non-radiative losses. Furthermore, we show that these resonators increase the measured fluorescence signal by enhancing the collection efficiency of emitted photons through a transparent substrate, in excellent agreement with numerical simulations. In practice, the interaction between a silicon nanoantenna and solid-state emitters is fully characterized by tuning the position of a fluorescent nanosphere, with nanometer precision, in the vicinity of an e-beam fabricated silicon disk using scanning-probe microscopy. The continuous monitoring of the emission intensity and excited-state lifetime of the molecules then provides a three-dimensional map of the LDOS [34-36]. By experimentally showing how subwavelength-sized Si nanoantennas manipulate and engineer spontaneous light emission with low ohmic losses, this study highlights the potential of dielectric resonators for nanoscale photon management.

## II. EXPERIMENTAL RESULTS AND DISCUSSION

### A. Principle of the experiment

To control the relative axial and lateral position of solid-state emitters with respect to a silicon nanoantenna, a 100 nm diameter fluorescent sphere (Invitrogen Red Fluospheres) was grafted at the extremity of a tapered optical fiber, which was mounted on a tuning fork to control its height using a sheer-force feedback loop. A Si nanoantenna was then scanned in three-dimensions using a piezoelectric stage (Fig. 1-a). Nanometer-scale silicon disks were fabricated by electron-beam lithography and reactive ion etching in a 105 nm thick amorphous Si film, deposited on a 1 mm thick silica substrate (see Appendix A). Fig. 1-b-c show scanning electron microscopy (SEM) images of two nanoantennas featuring base diameters of 170 ± 5 nm and 250 ± 5 nm, respectively. The etching process introduces a typical 10°-15° clearance angle between the base and the top of the fabricated silicon disks. In practice, the sample consisted of twenty arrays of one hundred Si nanoantennas with diameters ranging from 110 nm to 300 nm (by increments of 10 nm).

Choosing a film thickness close to 100 nm and tuning the nanoantenna diameters between 100 nm and 300 nm allowed us to redshift the resonance wavelength of the first order Mie modes (induced electric and magnetic dipoles) over the entire visible to near-infrared range [37-42]. This spectral evolution is evidenced in Fig. 1-d and Fig. 1-e, which present a darkfield optical microscopy image of the entire sample and scattering spectra of selected Si nanoantennas, respectively. When varying the diameter from 110 nm (upper left) to 200 nm (second row right), the scattered light increased in intensity and shifted from the blue to the red part of the spectrum. As shown in Fig. 1-e, larger resonators (lowest two rows) featured quadrupolar resonances in addition to their dipolar modes, leading to broader scattering spectra and larger cross-sections. The absorption and emission spectra of the fluorescent nanosphere were centered on 580 nm and 610 nm, respectively (see Fig. 2-a). In order to analyze Si nanoantennas in and out of resonance with the emitters, we considered two disk diameters of 170 nm and 250 nm. As shown on Fig. 2-a, the smaller nanoantenna featured a maximum scattering cross-section, corresponding to induced electric and magnetic dipoles [43], that overlapped with the maximum of the fluorescence spectrum of the nanosphere. On the other hand, the dipolar and quadrupolar resonances of the 250 nm diameter Si nanoantenna were centered on 550 nm and 710 nm, respectively, with a local minimum of the scattering cross-section around 610 nm. These spectral properties are in very good agreement with the theoretical scattering cross-sections shown in Fig. 2-b, obtained using Finite Difference Time Domain simulations (FDTD) with the disk dimensions measured in SEM, the dielectric constant of the amorphous Si film estimated in ellipsometry, and considering a 10 nm thick silica layer on the surface of the resonator due to the etching process [44] as well as an infinite silica substrate.

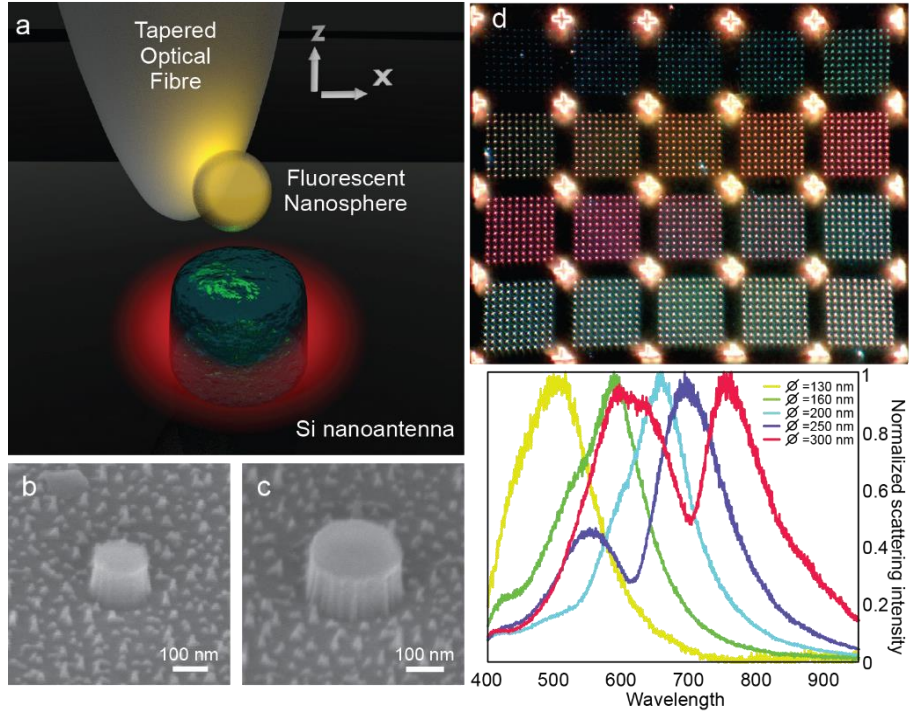

FIG. 1. (a) Sketch of the experimental study. SEM images of Si nanoantennas with base diameters of 170 nm (b) and 250 nm (c) (105 nm thickness). (d) Darkfield image of 20 arrays of 100 Si disks with diameters ranging from 110 nm (upper left) to 300 nm (lower right). The distance between two disks is 4 μm.

**B. Spontaneous emission enhancement and inhibition**

Near-field optical measurements were performed in epifluorescence, through the 1 mm silica substrate, with a 0.7 numerical aperture (NA) objective used both for the 560 nm pulsed excitation (Fianium SC450, 40 ps pulses, 10 MHz rate) and to collect the fluorescence signal at 610 nm on avalanche photodiodes PDM-R detector by Micro Photon Devices [45], based on a recent silicon avalanche photodiode technology developed for enhanced sensitivity in the near infrared [46], and combined with a Picoquant HydraHarp 400 acquisition board. The fluorescence lifetime and emission count-rate of the nanosphere were measured while scanning the Si nanoantenna in three dimensions with respect to the tapered optical fiber. The lifetime, τ, was estimated, for each position of the nanosphere, using time-correlated single photon counting and by fitting the decay curve with a single exponential decay [35, 36]. The measured decay rate, $\Gamma = 1/\tau$, therefore corresponds to an averaged value over the three orientations of the molecular transition dipoles and the spatial positions of the ~$10^3$ emitters in the 100 nm diameter sphere [34].

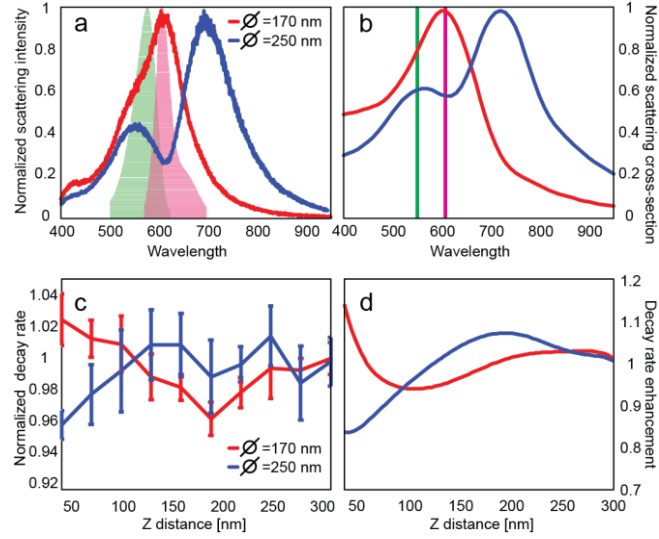

FIG. 2. Experimental (a) and simulated (b) scattering spectra of 170 nm (red solid line) and 250 nm (blue solid line) diameter Si nanoantennas. The faint green and purple spectra superimposed in (a) correspond to the absorption and emission spectra of the fluorescent nanosphere. The green and purple vertical lines in (b) denote the excitation laser line and maximum of the fluorescence emission. Experimental (c) and theoretical (d) evolution of the emission decay rate as a function of the axial distance between the fluorescent nanosphere and the 170 nm (red data points and solid line) or 250 nm (blue data points and solid line) diameter Si resonators.

Fig. 2-c shows the evolution of the excited-state decay rate of the fluorescent nanosphere when changing its axial position ($z$-axis) with respect to the 170 nm or 250 nm Si nanoantennas. The axial position corresponds to the distance between the top of the silicon disk and the bottom of the nanosphere when the latter is aligned with the center of the nanoantenna ($x = 0$). The change in $\Gamma$ was normalized with respect to the decay rate measured at $z = 300$ nm. We monitored oscillations of the decay rate similar to the ones observed when increasing the distance between a 2D interface and isolated solid-state emitters [47-50]. These damped oscillations are due to the constructive or destructive interference between the source dipoles in the nanosphere and the induced dipoles in the Si nanoantenna [48], whose phases depend on the emitter-resonator distance and on the polarizability of the antenna [18, 51]. Importantly, the oscillations in the vicinity of the "on resonance" and the "off resonance" antennas exhibited opposite behaviors: while the decay rate was increased close to the 170 nm diameter disk, it decreased in the vicinity of the 250 nm silicon nanoantenna. This measurement clearly highlights the interest of engineering the phase of the polarizability of dielectric resonators to influence spontaneous photon emission at the nanoscale. With plasmonic nanoantennas, the spectral dependence of the polarizability allows the control of emission directivity [51, 52] but high ohmic losses forbid the excited-state decay rate from being reduced in close proximity to the resonator, contrary to what is observed here with an "off resonance" 250 nm Si nanoantenna. Lifetime oscillations similar to Fig. 2-c can also be observed with planar films but the interaction with fluorescent emitters is one-dimensional, preventing the nanoscale engineering in three dimensions of inhibited or enhanced spontaneous photon emission.

To confirm the distance dependence of the emitter-resonator interaction, we simulated in Fig. 2-d the orientation-averaged decay rate enhancement for an emitter in the center of the polystyrene sphere as a function of the axial distance ($z$-axis) to the resonator. FDTD simulations were performed for three orthogonal electric dipole sources emitting at 610 nm, at the center of a 100 nm diameter sphere with a refractive index of 1.57 and with the silicon disk and substrate properties used for the theoretical spectra of Fig. 2-b (see appendix B). In particular, the shear-force feedback control of the tapered fiber and the grafting process of the polystyrene sphere produced a slight offset of the minimum axial separation between the fluorescent nanosphere and the nanoantenna surface, which could be estimated from the simulations. The $z$-dependence of the lifetime oscillations simulated in Fig. 2-d was in good agreement with the experimental data, for both resonator diameters, when considering a minimum of $35 \pm 5$ nm for the axial separation between the polystyrene sphere and the resonator. Furthermore, compared to the experimental conditions, the FDTD simulations neglected spatial averaging over the dimensions of the polystyrene sphere. As further illustrated in the following paragraphs, simulations considering a single orientation-averaged emitter provided a good agreement with experimentally measured spatial dependences of the fluorescence decay rate and intensity, although with higher amplitudes.

### C. Enhanced fluorescence collection efficiency

The nanoscale dimensions of broadband silicon nanoantennas allow for a local modulation of fluorescence emission in three dimensions. To illustrate this, we plot in Fig. 3 the measured and simulated evolution of the fluorescence count-rate when scanning laterally the fluorescent nanosphere over the 170 nm or 250 nm Si nanoantenna, with an estimated fixed axial separation of 120 nm with respect to the top of the resonator. We observe in Fig. 3-a that the collected fluorescence intensity was increased by 12 % when the sphere was above the "on resonance" silicon nanoantenna, compared to the count-rate of the sphere without the antenna (at the same axial separation from the silica substrate). In contrast, the fluorescence intensity was reduced by 40 % on top of the 250 nm diameter disk. When increasing the lateral separation between the nanosphere and the Si nanoantenna, the collected intensity was reduced by up to 38 % and 60 % for the smaller and larger disk, respectively, before recovering the intensity of the fluorescent sphere, without resonator, at distances larger than 700 nm.

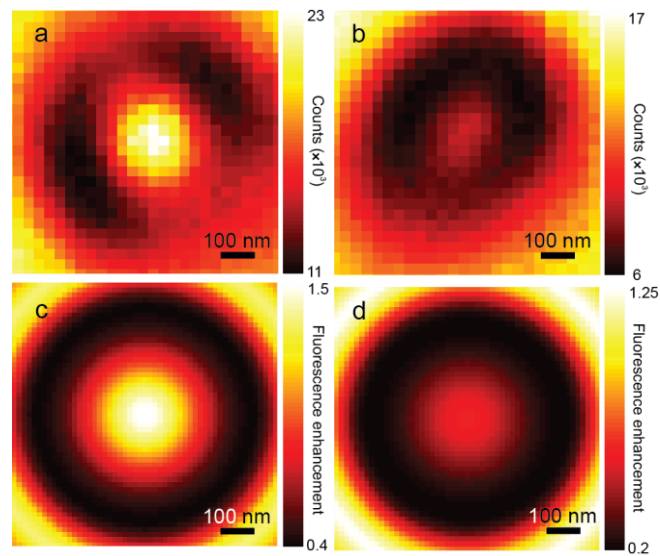

FIG. 3. Two-dimensional evolution of the measured (a,b) and simulated (c,d) collected fluorescence intensity in the vicinity of the 170 nm (a,c) and 250 nm (b,d) diameter nanoantennas. The centers of the two-dimensional maps correspond to the nanosphere centered on the dielectric disks.

In order to understand these results, we simulated two-dimensional fluorescence intensity maps. The simulations require the computation of the excitation intensity at 560 nm as well as the emitter quantum yield and photon collection efficiency at 610 nm. The collection efficiency was estimated by taking into account the silica substrate and the limited NA of the microscope objective while the initial quantum yield of the fluorescent emitter was set to 1. The simulated fluorescence intensities in the sample plane, given in Fig. 3-c-d, are in very good agreement with the experimental maps of Fig. 3-a-b. As mentioned above, the simulations did not consider the spatial distribution of fluorescent molecules in the sphere, which can account for the overestimated count-rate increase on the 170 nm diameter nanoantenna (the reduction with the "off resonance" antenna is in good agreement with the experimental data). However, the numerical simulations allow the separate analysis of the excitation intensity, the luminescence quantum yield and of the photon collection efficiency (Fig. 4-a-b-c). As shown in Fig. 4-a, a 170 nm diameter disk provided a 20 % enhancement of the excitation intensity but not the "off resonance" Si antenna. More importantly, interaction between the emitter and a resonant Si nanoantenna led to a 40 % increase in the collection efficiency (from 10% to 14%) compared to the case of the fluorescent sphere without a resonator (Fig. 4-b). Inversely, an "off resonance" antenna reduced the collected photon flux by 20 %. As shown in Fig. 4-d, the influence of the resonator size on the relative phase of the induced dipoles in the Si nanoantenna led to changes in the angular radiation pattern in the silica substrate and, thus, to a reduced collection efficiency for the "off resonance" case. On the other hand, the quantum yield of the emitter was similar in the vicinity of both antennas (Fig. 4-c) with a 10 % reduction because of the imaginary part of the dielectric constant of amorphous silicon. Finally, the reduced collected fluorescence intensity at distances of 200 nm – 300 nm from the resonators, seen in Fig. 3, was related both to a reduction of the excitation intensity and to a lower percentage of photons emitted in the silica substrate.

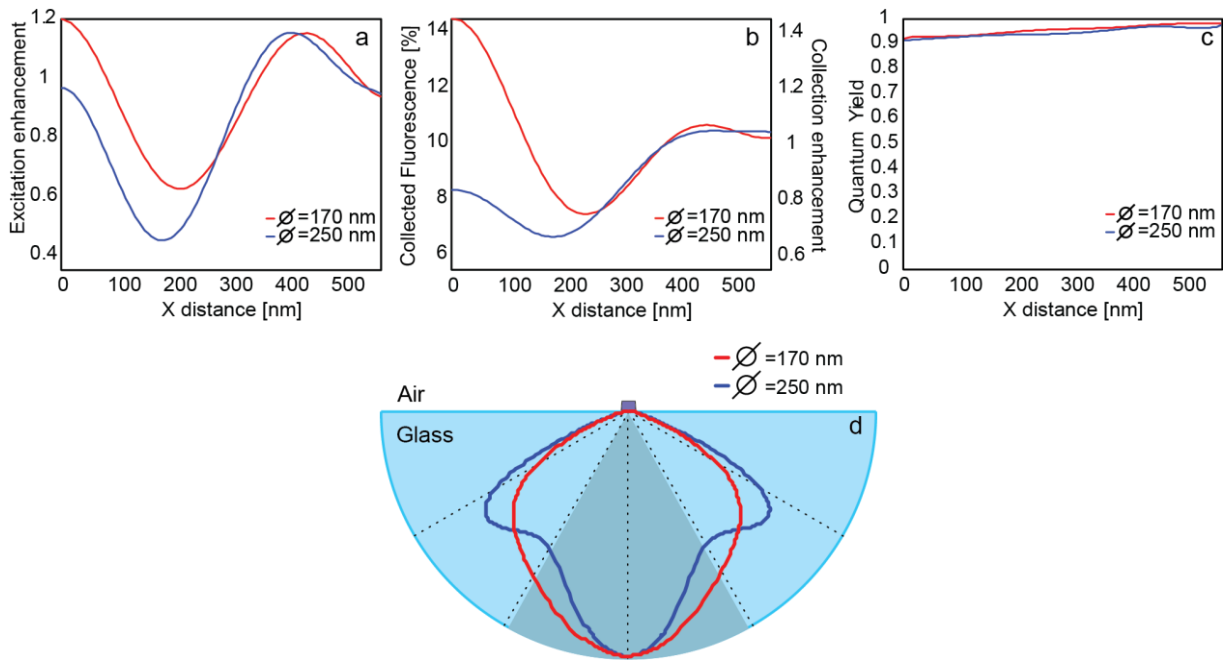

FIG. 4. Theoretical evolution of the excitation intensity enhancement in the center of the fluorescent sphere (a), of the fluorescence collection efficiency (b) and of the orientation-averaged quantum yield (c), as a function of the lateral distance between the nanosphere and the Si nanoantenna, with an axial separation of 120 nm above the 170 nm (red solid lines) or 250 nm (blue solid lines) diameter resonator. x = 0 corresponds to the nanosphere centered on the resonator. (d) Theoretical angular radiation pattern in the lower half-space when the fluorescent sphere is 120 nm above the 170 nm (purple solid line) or the 250 nm (blue solid line) silicon nanoantenna. The darker part highlights the numerical aperture of the microscope objective.

### D. Nanoscale enhancement of spontaneous emission

While Fig. 3 shows how silicon nanoantennas can modulate the collected fluorescence intensity, it does not demonstrate their ability to engineer the LDOS in three dimensions at the nanoscale. This is because, with a fluorescent sphere located at $z$ = 120 nm, the lifetime of the emitters was similar for both resonator sizes (see Fig. 2-c-d) and its fluctuations remained within the experimental error when scanning laterally the Si nanoantenna. We thus reduced the axial separation between the fluorescent nanosphere and the 170 nm diameter Si antenna down to 35 nm, which is the lowest value allowed by the shear-force feedback loop applied to the tapered optical fiber. At these short distances, a strong nanoscale modulation of the excited-state decay rate was observed (Fig. 5-a-b). In particular, both experimental and simulated values for the 2D spatial fluctuations of $\Gamma$ revealed an enhancement region centered on the nanoantenna with a typical size of 100 nm (Fig. 5-a-b). Furthermore, by radially averaging the spatial modulations of the fluorescence decay rate in Fig. 5-a, the line-scan of Fig. 5-c shows a 15 % increase of $\Gamma$ at the center of the resonator ($x$ = 0), which is in excellent agreement with the simulations provided in Fig. 5-d.

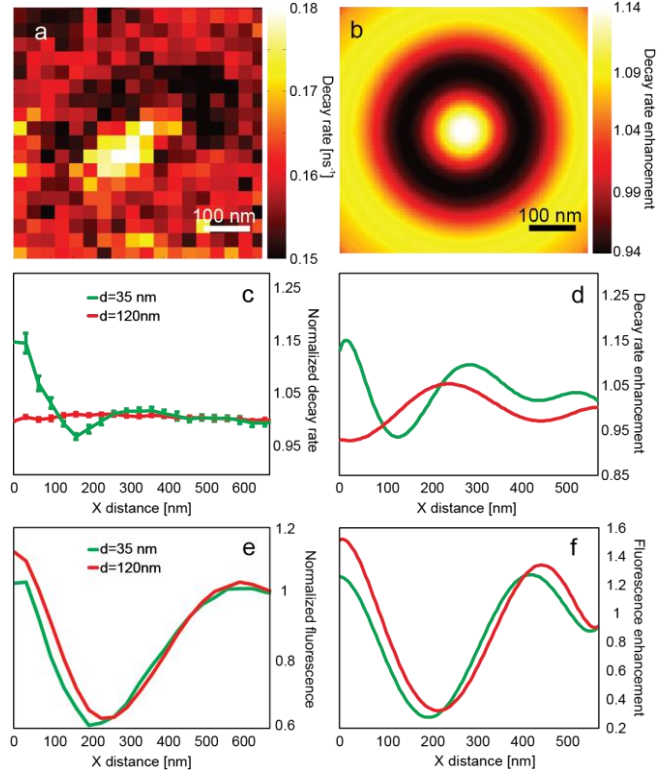

FIG. 5. 2D experimental (a) and simulated (b) maps of the total fluorescence decay rate in the sample plane, with an axial separation of 35 nm. Evolution of the experimental (c,e) and simulated (d,f) fluorescence decay rate (c,d) and intensity (e,f) as a function of the lateral distance between the 170 nm resonator and the fluorescent sphere for axial distances of 35 nm (green data points and solid lines) and 120 nm (red data points and solid lines). x = 0 corresponds to the nanosphere centered on the resonator.

As a reference, Fig. 5-c also shows the spatial evolution of the total decay rate when the axial separation between the sphere and the 170 nm nanoantenna was $z = 120$ nm. As already mentioned, the lifetime fluctuations were within the experimental error. Nevertheless, the fluorescence lifetime changes measured in Fig. 5-c featured significantly weaker fluctuations as compared to the simulations of Fig. 5-d. This is again due to the spatial averaging of the LDOS changes over the fluorescent sphere diameter, which was neglected in the simulations. Importantly, both experiments and simulations for $z = 35$ nm (Fig. 5-c-d) indicated a reduced LDOS at a lateral distance of $x \approx 120$ nm, demonstrating that the interference between the source dipoles and the dipoles induced in the Si nanoantenna led to enhanced and inhibited spontaneous emission in the sample plane similarly to what is observed along the $z$-axis in Fig. 2.

While changes in the axial separation between the sphere and the Si antenna strongly modified the total decay rate of the emitters, its influence on the measured fluorescence count rates was much weaker, as shown on Fig. 5-e-f. In particular, the solid red lines in Fig. 5-e-f correspond to the radially averaged data of Fig. 3-a-c. Both experimental data and simulations indicated similar damped spatial oscillations: the measured intensity was enhanced above the resonator but decreased at a lateral distance of about 200 nm before increasing again at larger spacings. As in Fig. 3, the simulated fluorescence count rate exhibited spatial fluctuations that are in good agreement with the experimental data, albeit with weaker amplitudes. Although lowering the position of the fluorescent sphere along the $z$-axis did not significantly influence the collected fluorescence signals shown in Fig. 5-e-f, numerical simulations indicate that it increased the photon collection efficiency but reduced the excitation enhancement and the fluorescence quantum yield (Fig. 6-a-b-c). In particular, reducing the axial distance of the fluorescent sphere above the resonator, from 120 nm to 35 nm, led to a two-fold increase in the collection efficiency enhancement (from 40 % to 80 %, Fig. 6-b) but decreased the excitation enhancement (from a 20 % gain to a 10 % reduction, Fig. 6-a) and reduced the quantum yield (from 90 % to 80 %, Fig. 6-c). The enhanced collection efficiency and decreased quantum yield are expected when increasing the coupling between emitters and a resonator that radiates efficiently into the silica substrate but features some ohmic losses. However, the reduced excitation intensity was more surprising. As shown in Fig. 6-d, the local intensity was reduced for a fluorescent sphere centered above the 170 nm Si disk, and the excitation field can only be significantly enhanced on the sides of the resonator. While a lower excitation field intensity should lead to reduced radiative decay rates for fluorescent molecules in the sphere (because of reciprocity) [53], we measured enhanced total decay rates in Fig. 5, close to the "on resonance" nanoantenna, because the emitter – resonator interaction was dominated by longitudinally oriented molecules (featuring a transition dipole perpendicular to the sample plane, see Fig. 6-e) while the simulated local field enhancement of Fig. 6-d was governed by transverse polarizations (in the sample plane).

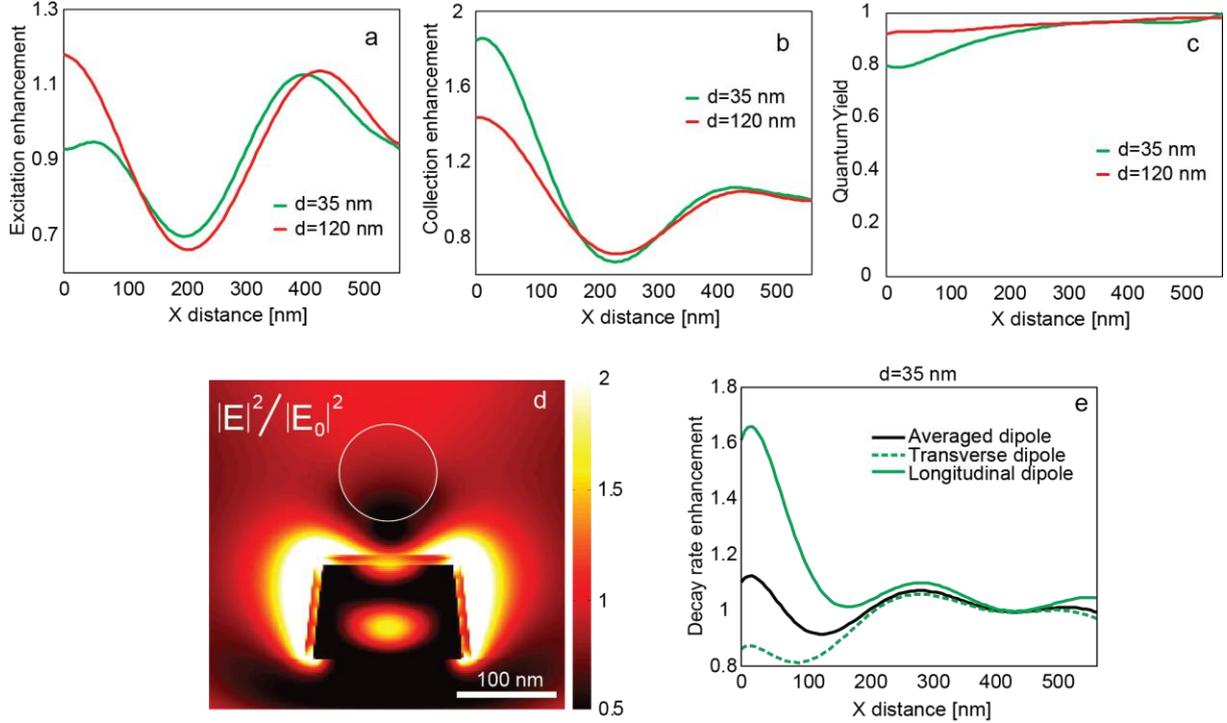

FIG. 6. Theoretical evolution of the excitation intensity enhancement in the center of the fluorescent sphere (a), of the fluorescence collection efficiency enhancement (b) and of the orientation-averaged quantum yield (c), as a function of the lateral distance between the nanosphere and the 170 nm diameter Si nanoantenna, with an axial separation of 35 nm (green solid line) or 120 nm (red solid line). (d) Spatial distribution of the electric field intensity, in a longitudinal plane containing the 170 nm diameter silicon antenna and the fluorescent sphere, for a 560 nm excitation wavelength (axial separation of 35 nm). (e) Simulated decay rate enhancement for a longitudinal (solid green line), a transverse (dashed green line) or three orientation-averaged dipoles (solid black line) in the center of the sphere as a function of the lateral distance between the nanosphere and the 170 nm diameter Si antenna, with an axial separation of 35 nm.

### III. Conclusion

Overall, the scanning probe technique used in this study has allowed us to characterize the resonant interaction between a fluorescent nanosphere and a silicon nanoantenna and to demonstrate inhibition and enhancement of spontaneous decay processes at the nanoscale and in three dimensions. Furthermore, good agreement between electrodynamics simulations and experimental data prove that the resonant behavior of nanometer-scale silicon particles allows them to modulate the emission and far-field collection of fluorescence photons and makes them excellent candidates for the design of optical nanoantennas with low ohmic losses. In particular, the simulated fluorescence intensity enhancements clearly show that the gain in measured photon count rates for the resonant 170 nm diameter Si nanoantenna is dominated by enhanced collection efficiency (by up to 85 %) and not by changes in excitation probability or, by reciprocity, of the local density of optical states (up to 15 % increase in total spontaneous decay rate). This result not only highlights the interest of dielectric antennas to control the directivity of emitted photons but also the need to improve the near-field coupling between emitter and resonator. The larger field confinements expected in coupled silicon particles [26, 29, 30, 32], associated with the top-down nanofabrication [52] or self-assembly strategies [6-8] developed for plasmonic nanoantennas, will allow the optimized interaction between solid-state light sources and dielectric-based resonators.

## ACKNOWLEDGEMENTS

The authors thank F. Bedu for assistance on sample fabrication. This work received funding from LABEX WIFI (Laboratory of Excellence within the French Program "Investments for the Future") under references ANR-10-LABX-24 and ANR-10-IDEX-0001-02 PSL*, and by Region Ile-de-France in the framework of DIM Nano-K. Work at Institut Fresnel was carried out with the support of the A*MIDEX project (No. ANR-11-IDEX-0001-02) funded by the Investissements d'Avenir French Government program and managed by the French National Research Agency (ANR). M. M. and M.F. G.-P gratefully acknowledge support from Fundacio Cellex and the Spanish Ministry of Economy and Competitiveness (SEV-2015-0522 grant). Nanofabrication processes were performed in the PLANETE cleanroom facility.
D. B. and M. M. contributed equally to this work.

## APPENDIX A: EXPERIMENTAL DETAILS

**Sample fabrication.** A thin film of amorphous silicon was obtained by e-beam evaporation of a solid source on a 1 mm thick UV-grade fused silica substrate in high vacuum. After chemical cleaning, the substrate was introduced in the growth chamber and underwent a smooth plasma cleaning before deposition of the silicon layer at room temperature. The substrates were then transferred in a tubular furnace for a 1h annealing step at 600 °C in vacuum to increase the density of the silicon thin film. The thickness (105 nm) and optical constants of the annealed films were determined ex situ by ellipsometry (n = 4.55 + 0.34$i$ at 610 nm).

After oxygen plasma cleaning to remove organic residues and increase the wettability of the silicon surface, the annealed amorphous film was covered with a 60 nm thick layer of e-beam resist (PMMA diluted at 2% in ethyl-lactate, AR-P 679.04 from ALL-RESIST, Germany) by spin-coating at 6000 rpm. After a soft baking process (10 min at 170 °C on a hot plate), a 30 nm thick conductive resist was spin-coated on the e-beam resist (SX AR-PC 5000/90.1 from ALL-RESIST) and baked 2 min at 85 °C. The e-beam lithography was then performed by a SEM-FEG system (PIONEER from RAITH). After exposure, the conductive resist was removed with deionized water and the sample was dried under nitrogen. The development was performed for 60 s in a commercial solution (AR 600-55 from ALL-RESIST, containing methyl isobutyl ketone (MIBK) and propan-2-ol (IPA)). The development was stopped in IPA and the sample dried under nitrogen. A 15 nm thick nickel mask was then evaporated under vacuum (Auto 306 tool from Edwards). Nickel was used as the mask material because it is resistant to fluorine attacks [54]. A lift-off process in acetone, using an ultrasonic bath, removed the e-beam resist and the excess of nickel in areas of the sample that were not irradiated with the electron beam.

The nickel mask pattern was transferred into the amorphous thin silicon film by reactive ion etching (MG200, Plassys, France). The unprotected areas were etched by the plasma of a gas mixture containing $SF_6$, $O_2$ and $CHF_3$ (respective fluxes of 20, 8 and 5 sccm), alternated with a pure $O_2$ plasma. Excited $SF_6$ efficiently etches silicon and the two key roles of the $CHF_3$ gas are to passivate the vertical feature walls [54] and to etch the silicon oxide that forms during the process on the very reactive amorphous silicon surface. This configuration allows a very good etching anisotropy and nearly vertical walls for the fabricated structures. The RF power used to excite the plasma was 120W for the gas mixture and 15W for pure $O_2$. This sequence was repeated as many times as required for the total etching of the thin silicon film. The nickel mask was removed by immersion in an aqueous $FeCl_3$ solution (3 min). Then the sample was rinsed by deionized water and dried under nitrogen flow.

**Scanning electron microscopy.** After all optical experiments were carried out, the sample featuring silicon resonators was covered by a 5 nm thick titanium layer and imaged with the GEMINI scanning electron microscope (1 nm resolution) of a Zeiss Auriga 60 FIB-SEM.

**Darkfield microscopy.** Darkfield images and spectra were measured in an inverted microscope (IX71, Olympus) coupled to a color CCD (Quicam, Roper) and to a fiber-coupled (50 μm core diameter) imaging spectrometer (Acton SP300 with Pixis 100 CCD detector, Princeton Instruments). White light from a 100 W halogen lamp was focused on the sample using a 0.8-0.92 NA darkfield condenser and the scattered light is collected, through the 1 mm thick silica substrate, with a 60× 0.7 NA objective. Darkfield spectra of individual silicon nanoantennas were measured by aligning a single disk in the confocal volume corresponding to the 50 μm fiber core of the imaging spectrometer with a 1 s acquisition time. A background spectrum was measured with the same acquisition time on an empty area of the sample. The background was subtracted to the measured spectra before correction according to the wavelength-dependent illumination and detection.

**Near-field optical measurements.** The sample was mounted on a sample-scanning inverted confocal microscope (Olympus, IX 71). A fluorescent polystyrene sphere (Invitrogen Red Fluospheres, 100 nm diameter) was grafted at the extremity of a tapered optical fiber (100 nm tip curvature) that was mounted on a tuning fork to control its height using a sheer-force feedback loop [35, 36]. Pulsed laser excitation at 560 nm (Fianium SC450, 40 ps pulses, 10 MHz rate) was focused on the sample, through the silica substrate, with a 0.7 NA 60× objective. Fluorescence photons were collected through the same objective and were separated from the excitation with a dichroic mirror and a high-pass filter ($\lambda > 580$ nm). Time-resolved photon detection was performed with a time correlated single photon counting system (PDM-R detector by Micro Photon Devices, combined with a Picoquant HydraHarp 400 acquisition board), which allows one to simultaneously map the fluorescence intensity and decay rate (instrument response function with a 190 ± 30 ps full-width at half maximum). The tapered optical fiber, supporting the fluorescent sphere, was held by shear force feedback at a constant distance of approximately 20 nm to the surface of a nanostructured sample while the latter was scanned (control electronics by RHK). The pixel size in the 2D intensity and decay rate maps is 30 nm × 30 nm with a 1 s acquisition time. Nanoscale drift of the experimental setup during the total measurement time of a 2D map (1000 s) explains the observed astigmatism in Fig. 3 and Fig. 5.

**Data analysis.** To estimate the data and error bars of Fig. 2-c, the evolution of the decay rate when changing the height of the fluorescent nanosphere was measured four (resp. three) consecutive times in the vicinity of the 170 nm (resp. 250 nm) diameter Si nanodisk. Each measurement consisted in a series of 10 successive acquisitions (10 s integration time), ranging from 35 nm to 300 nm. Fig. 7 shows typical fluorescence lifetime measurements close to the "on-resonance" and "off-resonance" nanoantennas or with a large axial distance of $z = 300$ nm. The experimental results presented in Fig. 2-c are the estimated mean decay rates when fitting these data with a single-exponential decay, while the error bars correspond to the 95 % confidence interval (plus or minus twice the standard error). The mean decay rates were normalized with respect to the mean value estimated at $z = 300$ nm.

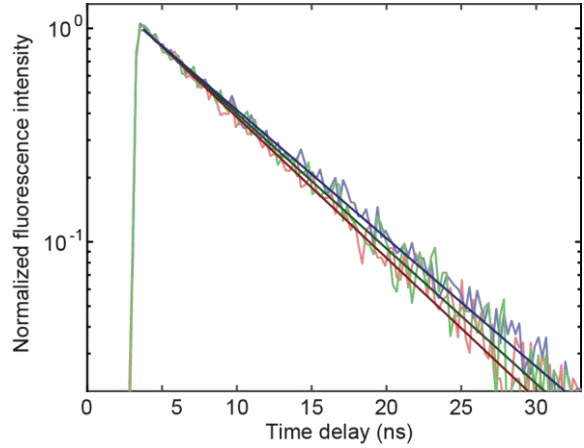

FIG. 7. Fluorescence intensity decay of a 100 nm nanosphere at an axial distance *z* = 35 nm from the 170 nm diameter (red data points, τ = 6.58 ± 0.06 ns) and the 250 nm diameter (blue data points, τ = 7.25 ± 0.05 ns) Si nanoantennas. The green data points provide the intensity decay measured with an axial distance of 300 nm from the 170 nm antenna (τ = 6.85 ± 0.08 ns). The lifetimes are estimated from single exponential fits (solid lines).

The data of Fig. 5-c and 5-e were estimated by radially averaging the decay rate and intensity maps of Fig. 3-a and 5-a. The center of the disk was first identified using the corresponding intensity map. All possible line-scans going from the center of the disk to the edge of the map were then plotted, resulting in a set of measurements of the emission decay rate and intensity as a function of the distance from the resonator center, in the sample plane. The experimental values presented in Fig. 5 are, for each distance, the mean decay rates and intensities, which were normalized with respect to the value estimated at *x* = 650 nm. The error bars in Fig. 5-c, for each estimated decay rate, correspond to the 95 % confidence interval.

## APPENDIX B: NUMERICAL SIMULATIONS

Numerical simulations were performed using an in-house Finite Difference Time Domain (FDTD) code [55]. The model took into account the nanoantenna dimensions estimated in SEM (105 nm thickness, 170 nm / 250 nm diameters), the dielectric constants of the amorphous silicon thin film measured by ellipsometry and an infinite silica substrate. A 10 nm thick layer of silicon dioxide (n=1.5) was added around the silicon resonator to account for the oxidation of the semiconductor matrix after the RIE treatment. The fluorescent molecules in the polystyrene sphere were modelled by three dipoles, oriented along the *x*, *y* and *z* axes, in the center of a 100 nm sphere with a dielectric index of 1.57. For simplicity, the tapered optical fiber was not taken into account. The simulations considered a volume spanning ± 2 μm in *x*, *y* and *z* around the nanostructures. The antennas were surrounded by air (optical dielectric index of 1). All six boundaries of the computational volume were terminated with convolutional-periodic matching layers to avoid parasitic unphysical reflections around the nanostructures. The non-uniform grid resolution varied from 25 nm, for areas at the periphery of the simulations, to 5 nm for the region in the immediate vicinity of the antennas.

Scattering spectra were simulated by impinging an unpolarized plane wave on the silicon nanoantenna from the upper air side of the sample. The scattered light was then collected in the silica lower half-space to account for the transmission configuration of the darkfield measurement.

Local excitation enhancement in the center of the sphere was estimated by impinging a plane wave propagating perpendicularly to the sample plane, from the silica substrate, on the sample. Changes in the total decay rate were obtained by computing the orientation-averaged changes in power dissipated by the three orthogonal dipoles in the center of the polystyrene sphere, when changing its *x*, *y* and *z* position with respect to the silicon antenna. The orientation-averaged radiative decay rate evolution, due to the resonator, was estimated by projecting the Poynting vector, produced by each of the three source dipoles, on the surface of the computation volume. Finally, the collection efficiency was calculated by integrating the projected Poynting vectors on the limited numerical aperture of the microscope objective (solid angle with a 56° width in the infinite silica substrate). For all calculations, we considered that the source electric dipoles have an intrinsic quantum yield of 1 and normalized the changes in local field distribution, as well as dissipated and radiated powers, with respect to a 100 nm polystyrene sphere with a given ($z$ + 105 nm) height above the silica substrate, without the silicon antenna. Simulations were performed for each axial (*z*-axis) and lateral (*x*-axis) position of the fluorescent nanosphere with respect to the center of the Si nanoantenna. The minimum considered axial distance (35 nm) is limited by the position of the fluorescent nanosphere on the tapered optical fiber and by the applied shear-force feedback loop during optical measurements. For shorter axial distances, strong variations of the excitation field and of the LDOS over the dimension of the nanosphere would require simulating the emission properties of source dipoles at different positions of the nanosphere and not simply in its center.

[†]valentina.krachmalnicoff@espci.fr
[‡]sebastien.bidault@espci.fr